\documentclass[aps,prl,twocolumn,superscriptaddress,amsmath,amssymb,showpacs]{revtex4}

\usepackage{graphicx}
\usepackage{dcolumn}
\usepackage{bm}

\begin{document}

\preprint{preprint}

\title{Structure and Magnetic Order in the NdFeAsO$_{1-x}$F$_x$  
Superconductor System}

\author{Y. Qiu}
\affiliation{NIST Center for Neutron Research, National Institute of Standards
and Technology, Gaithersburg, MD 20899, USA}
\affiliation{Department of Materials Science and Engineering, University of Maryland, College Park, MD 20742, USA}
\author{Wei Bao}
\email{wbao@lanl.gov}
\affiliation{Los Alamos National Laboratory, Los Alamos, NM 87545, USA}
\author{Q. Huang}
\author{T. Yildirim}
\affiliation{NIST Center for Neutron Research, National Institute of Standards
and Technology, Gaithersburg, MD 20899, USA}
\author{J. M. Simmons}
\author{M. A. Green}
\affiliation{NIST Center for Neutron Research, National Institute of Standards
and Technology, Gaithersburg, MD 20899, USA}
\affiliation{Department of Materials Science and Engineering, University of Maryland, College Park, MD 20742, USA}
\author{J.W. Lynn}
\affiliation{NIST Center for Neutron Research, National Institute of Standards
and Technology, Gaithersburg, MD 20899, USA}
\author{Y.C. Gasparovic}
\author{J. Li}
\affiliation{NIST Center for Neutron Research, National Institute of Standards
and Technology, Gaithersburg, MD 20899, USA}
\affiliation{Department of Materials Science and Engineering, University of Maryland, College Park, MD 20742, USA}
\author{T. Wu}
\author{G. Wu}
\author{X.H. Chen}
\affiliation{Hefei National Laboratory for Physical Science at Microscale and
 Department of Physics, University of Science and Technology of China,
 Hefei, Anhui 230026, China}


\begin{abstract}
The transition temperature $T_C\approx 26$ K of 
the recently discovered superconductor LaFeAsO$_{1-x}$F$_x$\cite{Kamihara2008}
 has been demonstrated to be extremely sensitive to the lanthanide ion, 
 reaching 55 K for the Sm containing oxypnictides\cite{A033603,A033790,A034234,A034283,A042053,A042105,A042582,A043727,A044290}.
 Therefore, it is important to determine 
  how the moment on the lanthanide affects the overall
  magnetism in these systems. Here we report a neutron diffraction study of the Nd oxypnictides.
  Long ranged antiferromagnetic order is apparent 
    in NdFeAsO below 1.96 K. Rietveld refinement shows that both Fe and Nd magnetic ordering are
required to describe the observed data with the staggered moment 1.55(4) $\mu_B$/Nd  and 
0.9(1) $\mu_B$/Fe at 0.3 K. The other structural properties
such as the tetragonal-orthorhombic distortion are found to be very similar to
those in LaFeAsO. Neither the magnetic ordering nor the structural distortion
occur in the superconducting sample NdFeAsO$_{0.80}$F$_{0.20}$ at any temperatures down to 1.5 K.
\end{abstract}



\maketitle

The surprising discovery of high $T_C$ superconductivity in cuprates
 two decades ago has shifted attention to laminar magnetic materials 
 for new high $T_C$ superconductors. New superconductors have been
  discovered since then in layered ruthenate\cite{214SC} and 
  triangular materials\cite{SC_Nb,H2OSC,Sc_cava}. While these 
  discoveries broke new ground for physics, their $T_C$'s are 
  not high. The recent discovery of high $T_C$ superconductors 
  in the quaternary Fe oxypnictides $Ln$FeAsO$_{1-x}$F$_x$ has 
  revitalized the field, and the
question naturally arises as to how the new family of
 iron-based superconductors compare to the cuprates. 
 Parent compounds of the new superconductors share a 
 similar electronic structure with all five $d$-orbitals 
 of the Fe contributing to a low density of states at the 
 Fermi level\cite{bs07,A030429,A031279,A033236,A033286,A033426,A041239,A042252}.
  This contrasts with cuprates in which parent compounds are 
  Mott insulators with well defined local magnetic moments\cite{pwa87}. 
  On the other hand, the electron and hole doping phase 
  diagram of the Fe oxypnictide 
  systems\cite{Kamihara2008,Wen2008,A034384,A042105} is 
  remarkably similar to that of the cuprates, for which 
  the high-$T_C$ superconductivity occurs when the 
  antiferromagnetic order of the parent compounds is 
  suppressed by doping. This similarity has inspired a 
  flurry of theoretical and experimental works. 

  A common phase diagram for these compounds has emerged in which the 
 stoichiometric parent compound shows a structural anomaly around 150 K, 
  below which spin-density-wave (SDW) antiferromagnetic ordering\cite{A033426} 
 appears, which is due to nesting Fermi surfaces that are dominated 
 by electronic states of Fe\cite{A030429}. 
 This SDW ordering is shown to break the degeneracy between the
 $d_{xz}$ and $d_{yz}$ orbitals of the Fe ion, consistent with the observed tetragonal-orthorhombic
 structural distortion\cite{A042252}. 
 Superconductivity only 
 occurs when this anomaly is suppressed, which can be achieved in 
  a number of ways such as fluorine doping on the oxygen 
  site\cite{Kamihara2008,Wen2008,A034384,A042105}. 
  Despite these common features, a wider picture as 
  to how the moment on the lanthanide affects the overall
   magnetism or why the superconducting transition temperature
    varies so greatly with different lanthanide ions is still unclear. 
  So far, the magnetic 
  structure has been determined only for LaFeAsO in 
  a system without magnetic rare-earth elements\cite{A040795}. 
  Thus, it is instrumental to establish whether or not this 
  is a general feature of the $Ln$FeAsO$_{1-x}$F$_x$ systems.

The NdFeAsO$_{1-x}$F$_x$ system is the first one to have 
$T_C\ge 50$ K\cite{A034234} and now shares the honor 
with the $Ln$=Pr, Sm, and Gd compounds\cite{A042053,A042105,A042582,A044290}.
 In this study, we choose NdFeAsO to represent the 
 non-superconducting members of the system, and 
 NdFeAsO$_{0.80}$F$_{0.20}$ the superconducting ones. 
 Polycrystalline samples of 2.6 g NdFeAsO and 7.6 g 
 NdFeAsO$_{0.80}$F$_{0.20}$ were synthesized using the
 solid state reaction. We measured resistivity of 
 both samples using the standard four-probe method 
 from pieces from the same batch of samples synthesized 
 for neutron diffraction experiments. The NdFeAsO 
 sample shows a strong anomaly at $T_S\sim 150$ K (Figure~1), 
 slightly higher than the previously reported value 
 of $\sim$145 K\cite{A034384}, which testifies to the 
 good sample stoichiometry since $T_S$ is known to 
 decrease with doping\cite{A034384}. The superconducting 
 transition temperature of the NdFeAsO$_{0.80}$F$_{0.20}$ 
 sample is $T_C\approx 50$ K.

\begin{figure}
\includegraphics[height=7cm]{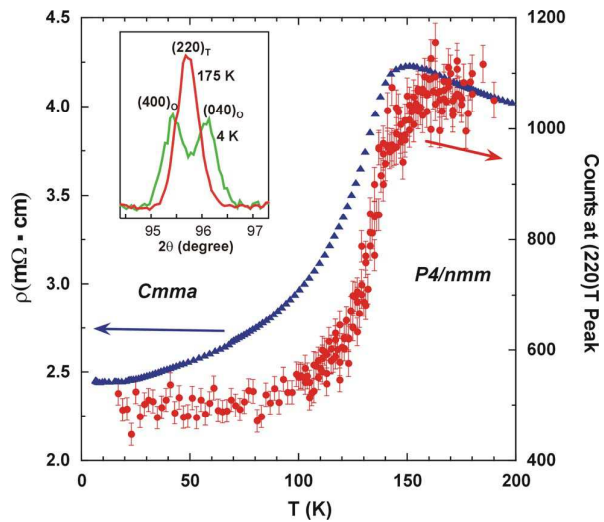} 
\caption{
(color online) Structural transition and its accompanying 
anomaly in resistivity in NdFeAsO. Both occur at 
$T_S\approx 150$ K. Above the transition, the 
crystal has the tetragonal $P4/nmm$ symmetry. 
Below $T_S$,
the crystal is distorted to have the orthorhombic 
$Cmma$ symmetry. 
The transition splits the tetragonal 
(220)$_T$ Bragg peak into orthorhombic 
(400)$_O$ and (040)$_O$ peaks as shown in the inset.
}
\label{fig1}
\end{figure}

Like in LaFeAsO, the resistivity anomaly is associated 
with a structure transition at $T_S$. Powder diffraction 
spectra of NdFeAsO measured at 175 and 4 K with neutrons 
of wavelength $\lambda=2.079\AA$, using the high 
resolution powder diffractometer BT1 at the 
NIST Center for Neutron Research (NCNR), are 
shown in Fig.~2(a) and (b) together with the 
refined profiles using the GSAS program\cite{gsas}. 
The high temperature structure is well described by 
the tetragonal ZrCuSiAs structure and the structure 
parameters at 175 K using space group $P4/nmm$ are 
listed in Table 1. Only small amounts of impurity 
phases, 1.5\% of Fe and less than 1\% of other 
impurities, were present in our NdFeAsO sample. 
The occupancy of all sites are within one standard
 deviation of the NdFeAsO sample stoichiometry, 
 therefore, the final refinement was performed with 
 the fixed stoichiometric occupancy. Below $T_S$, 
 NdFeAsO experiences an orthorhombic distortion. 
 The low temperature structure is well accounted 
 for by the space group $Cmma$ and the refinement
  at 4 K is shown in Fig.~2(b).

\begin{figure}
\includegraphics[height=14cm]{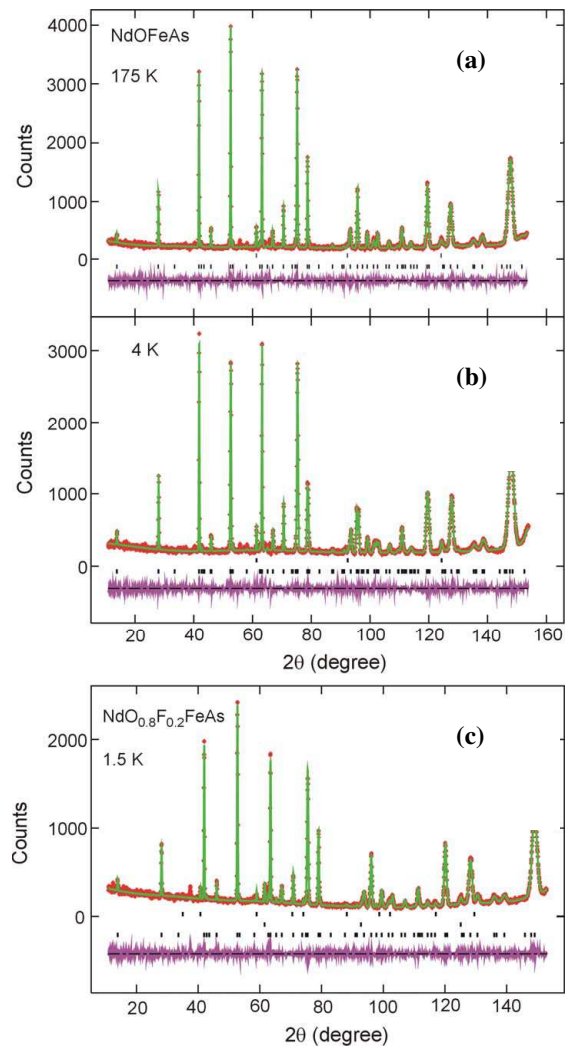} 
\caption{
(color online)
Neutron powder diffraction spectra of NdFeAsO at
(a) 175 and (b) 4 K, and (c) of NdFeAsO$_{0.80}$F$_{0.20}$ at 1.5 K.
While the spectra in (a) and (c) were refined by the space group $P4/nmm$,
the spectrum in (b) was by the space group $Cmma$.
}
\label{fig2}
\end{figure}

\begin{table*}
\caption{Refined structure parameters for NdFeAsO at 0.3 K and 175 K 
and NdFeAsO$_{0.8}$F$_{0.2}$ at 1.5K, respectively. Note that for the low-temperature 
structure of NdFeAsO, we also give the $\gamma$ angle  for the primitive cell of the Cmma space
group, which is about the same as  that of LaFeAsO. Similarly, the Fe-As distance is 2.40
\AA~ for both Nd and La compounds. The As-Fe-As angles are 111.2$^{\rm o}$ and 
108.82$^{\rm o}$ for NdFeAsO while the corresponding angles in LaFeAsO are 113.99$^{\rm o}$
and 107.06$^{\rm o}$, respectively. It seems that the closer the  
As-Fe-As angle is to the ideal tetrahedral-angle $\arccos(-\frac{1}{3})=109.47^{\rm o}$, the higher the T$_C$. 
}
\begin{center}
\begin{tabular}{|c|cccccccc|ccccc|ccccc|} \hline\hline
\multicolumn{1}{|c}{} &
\multicolumn{8}{|c}{NdFeAsO at 0.3 K } &
\multicolumn{5}{|c}{ NdFeAsO at 175 K} & 
\multicolumn{5}{|c|}{ NdFeAsO$_{0.8}$F$_{0.2}$ at 1.5 K } \\ 
\multicolumn{1}{|c}{} &
\multicolumn{8}{|c}{Space Group: Cmma} &
\multicolumn{5}{|c}{Space Group: P4/nmm} & 
\multicolumn{5}{|c|}{Space Group: P4/nmm} \\ 
\multicolumn{1}{|c}{} &
\multicolumn{8}{|l}{a=5.6159(1) \AA, b=5.5870(1) \AA, c=8.5570(2) \AA}&
\multicolumn{5}{|l}{a=b=3.9611(1) \AA, } & 
\multicolumn{5}{|l|}{a=b=3.9495(1) \AA} \\ 
\multicolumn{1}{|l}{} &
\multicolumn{8}{|l}{V=268.49 \AA$^{3}$ (Prim. Cell: a=b=3.9608 \AA, $\gamma=90.296^{\rm o}$)}&
\multicolumn{5}{|l}{c=8.5724(2) \AA, V=134.51 \AA$^{3}$} & 
\multicolumn{5}{|l|}{c=8.5370(3) \AA, V=133.16 \AA$^{3}$} \\ 
\multicolumn{1}{|l}{} &
\multicolumn{8}{|l}{$R_p=3.62\%,~wR_p=4.86\% $, $\chi^2=1.511$} &
\multicolumn{5}{|l}{$R_p=4.95\%,~wR_p=6.46\%$} & 
\multicolumn{5}{|l|}{$R_p=5.74\%,~wR_p=7.72\% $} \\ \hline
Atom&site &x &y &z &B(\AA$^{2}$) &M$_x$($\mu_B$) & M$_z$($\mu_B$)&M($\mu_B$) &
site &x &y &z &B(\AA$^{2}$) & 
site &x &y &z &B(\AA$^{2}$) \\ \hline 
Nd & 4g & 0 &$\frac{1}{4}$ & 0.1389(2) & 0.06(5) & 1.22(7) & 0.96(9) & 1.55(4) &
2c  & $\frac{1}{4}$ &$\frac{1}{4}$ & 0.1393(3) & 0.29(6) &
2c  & $\frac{1}{4}$ &$\frac{1}{4}$ & 0.1421(4) & 0.54(5) \\
Fe & 4b & $\frac{1}{4}$ & 0 & $\frac{1}{2}$ & 0.68(4) &  0.9(1) & 0 & 0.9(1) &
2b  & $\frac{3}{4}$ &$\frac{1}{4}$ & $\frac{1}{2} $& 0.61(5) &
2b  & $\frac{3}{4}$ &$\frac{1}{4}$ & $\frac{1}{2} $& 0.12(4) \\
As & 4g & 0 &$\frac{1}{4}$ & 0.6584(4) & 0.97(8) &  &  &  &
2c  & $\frac{1}{4}$ &$\frac{1}{4}$ & 0.6580(4) & 1.00(8) &
2c  & $\frac{1}{4}$ &$\frac{1}{4}$ & 0.6599(4) & 0.54(5) \\
O/F & 4a & $\frac{1}{4}$ & 0 & $ 0 $ & 0.56(9) &    &   &   &
2a  & $\frac{3}{4}$ &$\frac{1}{4}$ & 0   & 0.68(7) &
2a  & $\frac{3}{4}$ &$\frac{1}{4}$ & 0   & 0.12(4) \\ \hline\hline
\end{tabular}
\end{center}
\end{table*}

The orthorhombic distortion doubles the unit cell, 
which is approximately 
$(\sqrt{2}a+\varepsilon)\times(\sqrt{2}a-\varepsilon)\times c$ 
in terms of the tetragonal unit cell. 
This splits the (220)$_T$ Bragg peak of 
the tetragonal structure into nonequivalent 
(400)$_O$ and (040)$_O$ Bragg peaks of the 
orthorhombic structure, see inset to Fig.~1. 
In terms of primitive cell parameters (Table~1), the distorted structure
has  $\gamma=90.296^{\rm o}$, which is almost identical to that of 
LaFeAsO\cite{A040795}, indicating Nd has almost no effect on the structural distortion. 
This is consistent with a recent theoretical study where it was shown that the structural
distortion is due to the ordering of the Fe $d_{xz},d_{yz}$ orbitals in the SDW 
structure\cite{A042252}.
To establish the relation between the structural 
transition and the resistivity anomaly for NdFeAsO,
 the intensity at the peak position of (220)$_T$ was 
 measured as a function of temperature. As shown in Fig.~1, 
 it is obvious that the anomaly in resistivity is caused 
 by the structural phase transition at $T_S\approx 150$ K, 
 like previously discovered in LaFeAsO\cite{A040795}.

There is no structural transition in our 
superconducting NdFeAsO$_{0.80}$F$_{0.20}$ sample. 
The tetragonal ZrCuSiAs structure well describes the observed powder 
diffraction spectra down to 1.5 K. Fig.~2(c) shows the spectrum 
at 1.5 K together with the refined profile. Only small amounts 
of impurity phases, 1.5\% of Fe and 4.4\% of NdAs, 
were present in the sample. Refined structure parameters 
at 1.5 are listed in Table~1. 
The lattice of the NdFeAsO$_{1-x}$F$_x$ system is more 
compacted than that of the LaFeAsO$_{1-x}$F$_x$ 
system\cite{A040795,A051062}, and the $T_C$ of the 
former is higher than $T_C$ of the latter. 
However, the structural transition 
temperature $T_S$ of the two systems remains the same.

\begin{figure}
\includegraphics[height=5.5cm]{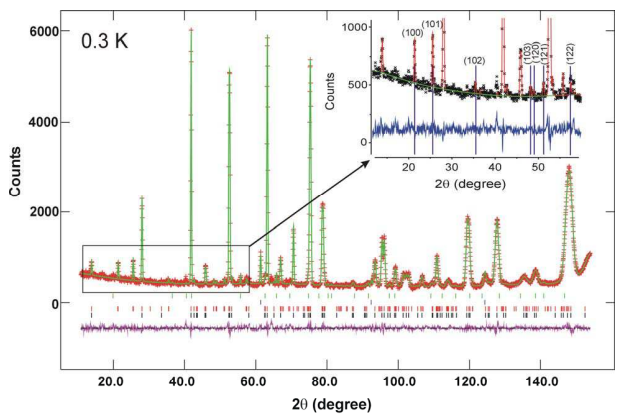} 
\caption{
(color online)
Neutron powder diffraction spectrum at 0.3 K.
The small angle part is magnified in the inset, 
and magnetic Bragg peaks which are absent at 4 K 
are marked by the vertical lines and are indexed 
using the orthorhombic $Cmma$ unit cell.
}
\label{fig3}
\end{figure}

Despite the similarity with the structural transition in 
NdFeAsO and LaFeAsO, we did not observed the SDW order of the type observed in LaFeAsO with the magnetic wavevector (1/2,1/2,1/2)$_T$ \cite{A040795} in either our NdFeAsO sample down to 1.5 K or the  NdFeAsO$_{0.80}$F$_{0.20}$ sample down to 6 K, using BT1 or higher flux triple-axis spectrometer BT9 or BT7 at the NCNR. Our measurements at BT9 and BT7 set the upper limit for the
staggered magnetic moment of the LaFeAsO-like SDW order at below 0.17 $\mu_B$ per Fe at 30 K for NdFeAsO and below 0.08 $\mu_B$ per Fe at 45 K for NdFeAsO$_{0.80}$F$_{0.20}$. 
However, the long ranged magnetic order is apparent below 2 K. 
In Fig.~3, neutron powder diffraction spectrum 
measured at BT1 at 0.3 K for the NdFeAsO sample 
in a $^3$He cryostat is shown. When compared to 
the spectrum taken at 4 K in Fig.~2(b), additional 
magnetic Bragg reflections are visible and marked 
by the vertical lines in the inset. In Fig.~4, the
temperature dependence of the magnetic Bragg peak 
(100)$_O$, measured at BT7, is shown. The solid line 
represents the mean-field theoretical fit for the 
squared magnetic order-parameter and the N\'{e}el 
temperature is determined as $T_N=1.96(3)$ K.

\begin{figure}
\includegraphics[height=6cm]{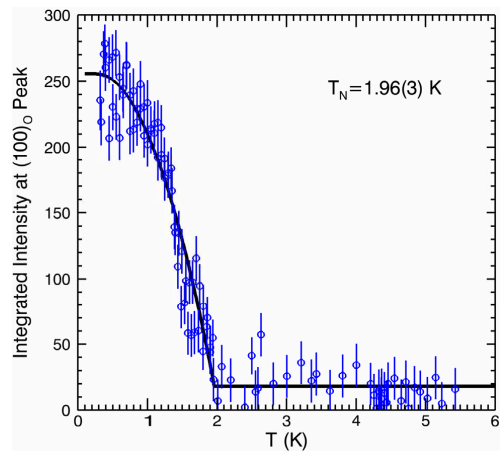} 
\caption{
(color online)
The intensity of magnetic Bragg peak (100)$_O$ 
as a function of temperature.
}
\label{fig4}
\end{figure}

The magnetic wave-vector (1/2,1/2,0)$_T$ of the low temperature 
antiferromagnetic order in NdFeAsO is consistent with the 
orthorhombic crystal structure below $T_S$.
Therefore, there is no further crystal symmetry breaking 
below $T_N$. The parallel alignment of magnetic moments 
along the $c$-axis demanded by the magnetic wave-vector 
in NdFeAsO differs from the antiferromagnetic one in LaFeAsO, 
which further doubles the unit cell along the $c$-axis. 
The low $T_N=1.96(3)$ K indicates the important role 
played by Nd in the antiferromagnetic transition, as 
rare-earth magnetic ions often order at low temperature, 
such as in the high $T_C$ cuprate Nd$_{2-x}$Ce$_x$CuO$_4$ 
system below $\sim$1.2 K\cite{jeff90,nd2cuo4}.
However, a magnetic model with Nd ions alone cannot account 
for the magnetic diffraction pattern, see Fig.~5. 
We solved the magnetic order using a combined Nd and 
Fe antiferromagnetic structure. 
Comparison of the Rietveld refinement fits to magnetic models with
different moment directions and the directions of
the ferro and antiferro-alignment of the moments clearly shows significantly better result for the model shown in Fig.~5
The resulting crystalline and magnetic refinement is 
shown in Fig.~3 and the structure and magnetic parameters 
at 0.3 K are listed in Table 1. The Fe moments are orientated along the longer of the two axis, $a$, the direction where Nd also has a component. While the antiferromagnetic alignment for Nd is along the $b$-axis, it is along the $a$-axis for the Fe moments, consistent with  previous
first-principles calculations\cite{A042252}.
  The total staggered
 magnetic moments are 1.55(4) $\mu_B$ per 
 Nd and 0.9(1) $\mu_B$ per Fe at 0.3 K.
  
\begin{figure}
\includegraphics[height=5.cm]{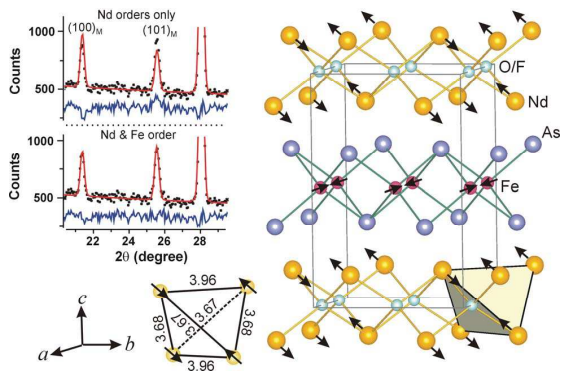} 
\caption{
(color online)
Magnetic structure of NdFeAsO below 
$T_N=1.96$ K in the orthorhombic unit cell. 
The Nd tetrahedron structure element is also shown.
A comparison of the Rietveld refinement fits 
  using a Nd only model as compared with a Nd/Fe 
  coupled model is included to highlight the necessity 
  for the latter to fully describe the scattering.}
\label{fig5}
\end{figure}

In conclusion, we presented a detailed study of the structural and magnetic properties of 
NdFeAsO and the doped 50~K superconductor NdFeAsO$_{0.80}$F$_{0.20}$. We clearly show that NdFeAsO has very
similar structural properties to LaFeAsO system, and the main differences show up at 
very low temperatures
where the Nd and Fe moments are ordered. 
We were not able to observe the Fe SDW ordering, suggesting that the Fe moment in NdFeAsO should
be much smaller than that in LaFeAsO. In fact, all-electron first-principles calculations,
similar to one in Ref.~[\onlinecite{A042252}] but using the LaFeAsO structure with the 
lattice parameters of NdFeAsO,  show that the
Fe moment is reduced from $0.48 \mu_B$ for La to $0.30 \mu_B$ for Nd. 
Since the Bragg intensity is
proportional to the square of the ordered moment, this suggests about $1/3$ intensity
decrease in NdFeAsO compared to the La-system, providing a possible explanation
why we did not observe the Fe SDW ordering in NdFeAsO.
From our results, it is tempting to conclude that the reason for the higher
T$_c$ in NdFeAsO is the different structure parameters due to smaller lanthanide ion (Table I)
and not due to other effects. 

Work at LANL is supported by U.S. DOE, 
at USTC by the Natural Science Foundation of China, 
Ministry of Science and Technology of 
China (973 Project No: 2006CB601001) 
and by National Basic Research Program 
of China (2006CB922005).

Note added: After the completion of our work, an independent neutron diffraction study on NdFeAsO$_{1-x}$F$_x$ was posted\cite{A061450}. 
Consistent with our observations, Bos et al.\ also only observed magnetic transition at very low temperature. However, they did not determine the N\'{e}el temperature, and their conclusion of the antiferromagnetic order formed by the Nd moments alone has been shown in our work as inadequate.


\end{document}